%% file: main_TBD.tex
\documentclass[nonacm,sigplan]{acmart}

\usepackage[normalem]{ulem}

\input{setup/packages}

\input{setup/macros}

\usepackage{enumitem}

\definecolor{codegreen}{rgb}{0,0.6,0}
\definecolor{codegray}{rgb}{0.5,0.5,0.5}
\definecolor{codepurple}{rgb}{0.58,0,0.82}
\definecolor{backcolour}{rgb}{0.95,0.95,0.92}

\newcommand{\circled}[1]{\tikz[baseline=(myanchor.base)] \node[circle,fill=.,inner sep=1pt] (myanchor) {\color{-.}\bfseries\footnotesize #1};}

\newcommand{\eat}[1]{}

\lstdefinestyle{mystyle}{
    backgroundcolor=\color{backcolour},   
    commentstyle=\color{codegreen},
    keywordstyle=\color{magenta},
    numberstyle=\tiny\color{codegray},
    stringstyle=\color{codepurple},
    basicstyle=\scriptsize, 
    breakatwhitespace=false,         
    breaklines=true,                 
    captionpos=b,                    
    keepspaces=true,                 
    numbers=left,                    
    numbersep=5pt,                  
    showspaces=false,                
    showstringspaces=false,
    showtabs=false,                  
    tabsize=1
}
\lstdefinelanguage
   [x64]{Assembler}     
   [x86masm]{Assembler} 
   {morekeywords={cmpq,movq,popq,pushq,subq,addq,retq,movslq,%
                  rax,rdx,rcx,rbx,rsi,rdi,rsp,rbp, %
                  r8,r8d,r8w,r8b,r9,r9d,r9w,r9b, %
                  r10,r10d,r10w,r10b,r11,r11d,r11w,r11b, %
                  r12,r12d,r12w,r12b,r13,r13d,r13w,r13b, %
                  r14,r14d,r14w,r14b,r15,r15d,r15w,r15b}} 

\begin{document}

\title{\sys: Protecting against Side-Channels Attacks\\ using Self-Destructing Enclaves}
\thanks{`Quan' represents quantum. In a quantum system, the information will be destroyed when attacked. Our system has the same property.}
%
\author{Shujie Cui}
\affiliation{%
  \institution{Monash University}
  \country{Australia}
}

\author{Haohua Li}
\affiliation{%
  \institution{Monash University}
  \country{Australia}
}

\author{Yuanhong Li}
\affiliation{%
  \institution{Monash University}
  \country{Australia}
}

\author{Zhi Zhang}
\affiliation{%
  \institution{University of Western Australia}
  \country{Australia}
}
\author{Lluís Vilanova}
\affiliation{%
  \institution{Imperial College London}
  \country{United Kingdom}
}
\author{Peter Pietzuch}
\affiliation{%
  \institution{Imperial College London}
  \country{United Kingdom}
}


\begin{abstract}
Trusted Execution Environments (TEEs) allow user processes to create \emph{enclaves} that protect security-sensitive computation against access from the OS kernel and the hypervisor. Recent work has shown that TEEs are vulnerable to side-channel attacks that allow an adversary to learn secrets shielded in enclaves. The majority of such attacks trigger exceptions or interrupts to trace the control or data flow of enclave execution.

  We propose \emph{\sys}, a system that protects enclaves from side-channel attacks that interrupt enclave execution. The main idea behind \sys is to strengthen resource isolation by creating an interrupt-free environment on a dedicated CPU core for running enclaves in which enclaves terminate when interrupts occur. \sys avoids interrupts by exploiting the tickless scheduling mode supported by recent OS kernels. \sys then uses the \emph{save area}~(SA) of the enclave, which is used by the hardware to support interrupt handling, as a second stack. Through an LLVM-based compiler pass, \sys modifies enclave instructions to store/load memory references, such as function frame base addresses, to/from the SA. When an interrupt occurs, the hardware overwrites the data in the SA with CPU state, thus ensuring that enclave execution fails. Our evaluation shows that \sys significantly raises the bar for interrupt-based attacks with practical overhead. 
\end{abstract}

\maketitle
\pagestyle{plain} 

\input{sections/intro}
\input{sections/background}

\input{sections/overview}

\input{sections/details}

\input{sections/implementation}

\input{sections/evaluation}
\input{sections/relatedwork}

\input{sections/conclusion}


\bibliographystyle{plain}
\bibliography{bibliography}

\end{document}

%% file: setup/packages.tex
\usepackage{tikz}
\usepackage{amsmath}

\usepackage{filecontents}

\usepackage{graphicx}
\usepackage{textcomp}
\usepackage{xcolor}
\usepackage{footnote}
\usepackage{booktabs}
\usepackage{times}

\usepackage{pgfplots}
\usepackage[linesnumbered,ruled,vlined]{algorithm2e}

\usepackage[hang,flushmargin]{footmisc}
\usepackage{marvosym}
\usepackage{diagbox}
\usepackage{caption}
\usepackage{subcaption}
\usepackage{hyperref}
\usepackage{todonotes}
\usepackage{multirow}
\usepackage{soul} 
\usepackage{listings}
\usepackage[T1]{fontenc}
\usepackage{beramono}

\usepackage{pifont}
\usepackage{setspace}


%% file: setup/macros.tex
\newcommand{\sys}{{QuanShield}\xspace}

\usepackage{xspace}
\newcommand{\etal}{et al.\xspace}

\newcommand{\ie}{i.e.,\xspace}
\newcommand{\eg}{e.g.,\xspace}

\mathchardef\mhyphen="2D

\newcommand{\code}[1]{\texttt{\small{}#1}\xspace}
\newcommand{\note}[1]{\noindent{\color{red}\textbf{#1}}}

\SetCommentSty{mycommfont}

\SetKwInput{KwInput}{Input}                
\SetKwInput{KwOutput}{Output}              

\newlength{\gapspace}

\renewcommand{\todo}[1]{\textbf{\color{red}[\textsc{todo:} #1]}}
\newcommand{\gap}[1][0cm]{\setlength{\gapspace}{#1}\vspace{.5\gapspace}\textbf{\color{red}[...]}\vspace{.5\gapspace}\xspace}

\renewcommand{\note}[1]{\noindent{\color{red}\textbf{#1}}}
\newcommand{\noteprp}[1]{\note{\color{orange}[\textsc{prp:} #1]}}

\newcommand{\tinyskip}{\vspace{3pt}}

\newcommand{\mypar}[1]{\tinyskip\noindent\textbf{#1.}\xspace}
\newcommand{\myparr}[1]{\tinyskip\noindent\textbf{#1}\xspace}

%% file: sections/intro.tex

\section{Introduction}
\label{sec:intro}

A trusted execution environment (TEE) is a secure and isolated environment within a computer system or a microprocessor that provides confidentiality and integrity for executing sensitive code and protecting sensitive data. It is designed to protect against various threats, such as malware and unauthorised access. At the hardware level, TEEs make use of secure \emph{enclaves}, which are separate regions of the memory created using specialised hardware features such as secure processors or trusted platform modules (TPMs). Different hardware architectures and platforms may have different names or implementations for TEEs. For example, Intel's Software Guard Extensions (SGX)~\cite{sgx14}, AMD SEV~\cite{SEV}, and ARM TrustZone~\cite{Trustzone} are three well-known TEE implementations used in different devices. 

However, the isolation provided by TEEs is not sufficient to protect the enclave. Regardless of the implementation, enclaves still share resources with untrusted code, \eg CPU caches, buffers, and branch prediction units. Moreover, most of the resources are managed by the underlying OS or hypervisor. This reliance on shared resources and the untrusted OS or hypervisor makes enclaves susceptible to \emph{side-channel attacks}~(SCAs). Recent work has extensively exploited side channels to infer secrets from enclaves, especially for Intel SGX~\cite{cacheout, WangCPZWBTG17, SchwarzWGMM20, memjam, ridl, Fallout, ZombieLoad, foreshadow}, ARM TrustZone~\cite{ChoZKPL0DA18, TruSense, TruSpy, ARMageddon}, and AMD SEV~\cite{LiZWLC21, CIPHERLEAKS, Crossline, LiZWLC21, Morbitzer0H19, LiZLS19, WernerMAPM19, Severed}.

A typical SCA triggers exceptions or interrupts during the execution of enclaves. TEEs rely on the untrusted OS to process exceptions or interrupts. Whenever the enclave is interrupted, it has to exit temporarily and give up the CPU to the kernel to process interrupts. Before transferring the CPU to the kernel, the hardware saves the CPU state, \ie its execution context, in a Secure Area (SA) in the memory and clears the registers. Whereas, the data in most of other resources, such as cache and on-chip buffers, are not cleaned and could be exploited by attackers. With tools such as SGX-Step~\cite{vanbulck2017sgxstep}, SEV-step~\cite{wilke2023sevstep}, and Load-step~\cite{Loadstep}, an adversary can control hardware interrupt controllers and single-step the execution. 

Existing defences against SCAs are either hardware- or software-based: (i)~\emph{hardware-based} defences propose new hardware mechanisms to eliminate SCAs, \eg through cache partitioning~\cite{CATalyst, ZhouRZ16, COLORIS} or self-paging~\cite{OrenbachBS20, self-paging}. They are effective but unsupported by current hardware; (ii)~\emph{software-based} solutions are either proactive or reactive: \emph{proactive} approaches normalise or randomise the memory access patterns~\cite{Strackxabs, drsgx, OBFUSCURO, Klotski}; \emph{reactive} approaches detect ongoing SCAs during the execution~\cite{ChenZRZ17, Cloak, chen2018, varys, Hyperspace}.

Proactive approaches generally impose a large performance overhead due to code transformations, especially those for protecting Intel SGX. For example, to protect SGX enclaves from page-fault-based SCAs, the technique proposed by Shinde \etal~\cite{ShindeCNS16} slows down execution by over $3000\times$; SGX-shield~\cite{SeoLKSSHK17} randomises page access patterns and enlarges enclaves by 3.7$\times$. Defences based on detection incur less overhead but suffer from false negatives, where the attacker manages to interrupt the enclave and learn secrets without being detected. For example, Varys~\cite{varys} periodically detects enclave exits and aborts the SGX enclave if it exits unexpectedly often. To ensure an acceptable performance overhead, Varys cannot detect infrequent exits. With SGX-Step~\cite{vanbulck2017sgxstep}, an adversary can interrupt an enclave multiple times while staying below Varys' detection threshold. 

We describe \textbf{\sys}, a new detection-based system that protects enclaves from interrupts-based SCAs. The main idea behind \sys is to strengthen the isolation by executing each enclave thread on a dedicated CPU in an interrupt-free environment. 
If an interrupt occurs regardless, it must have been triggered by an attacker. To detect attacks, \sys changes the enclave memory layout in a way that the hardware terminates the execution when an interrupt occurs. In contrast to existing detection-based defences~\cite{varys, Shih0KP17, ChenZRZ17}, \sys does not rely on detection code to count interrupts and thus does not require thresholds, which is costly and can be undermined by attackers. \sys thus achieves a lower performance overhead and fewer false negatives. 

\mypar{Interrupt-free execution} \sys ensures that each enclave thread executes on a CPU core that does not handle interrupts. This can be achieved with any recent OS kernels by operating the scheduler in \emph{tickless} mode. Tickless mode is a power-saving feature that only delivers timer interrupts as required. \sys disables all local timer ticks because it only executes a single hardware thread that is used to multiplex multiple user-level threads; other types of interrupts are delivered to CPU cores that do not execute the enclave.


\mypar{Attack detection} \sys exploits the observation that the SA is unused unless the enclave thread is attacked. To ensure enclave execution fails in such cases, \sys uses the SA as a \emph{second stack} to store memory references, such as each function's frame base address, and instruments the code to load memory references from the SA. 
After an interrupt has occurred, the memory references stored in the SA are overwritten with saved CPU state by the hardware. 
Modification of memory references violates invariants on memory accesses enforced by the CPU, and accessing such addresses fails enclave execution, thus preventing an attack. This approach does not require the execution of additional detection operations, thus making \sys more efficient than existing defences. 

\mypar{Performance and security evaluation} We take Intel SGX as an example to show the specific design and performance. \sys uses an LLVM compiler pass to instrument load/store memory references into/from the SA, without requiring manual source code changes. We measure the security guarantee of \sys in two ways: the crash rate when attacked and the response delay to attacks. We measure the latter with the number of interrupts an attack can send to the enclave before it crashes, which determines the amount of information an attacker could learn. We implement a prototype of \sys that provides block-granularity protection, \ie if a block is attacked, the enclave crashes before executing the next block. \sys achieves that by storing each function's frame base address in the SA and reloading it at the beginning of each block. In this case, the test results show that \sys terminates the enclave with a 100\% rate and the response delay matches with the block size. For instance, when the block contains less than 9 instructions on average, \sys terminates the enclave within 7 interrupts. We evaluate \sys{}'s performance overhead and compare it with a baseline in which the enclave executes in an interrupt-free environment without SCA protection. Our results show that \sys{}'s overhead is less than 15\% when the enclave has less than 3000 blocks. We also compare the performance of \sys with Varys since it is also a pure software-based detection solution against interrupt-based attacks. When Varys and \sys provide comparable security, \sys is up to $2.8\times$ faster than Varys. 

%



%% file: sections/background.tex

%
\section{Background}
\label{sec:background}

\subsection{Trusted Execution Environments}

A Trusted Execution Environment (TEE) leverages a combination of hardware and software mechanisms to ensure security. 
At the hardware level, secure processors, like Intel Software Guard Extensions (SGX) or AMD Secure Encrypted Virtualisation-Secure Nester Paging (SEV-SNP), provide hardware-based memory encryption and integrity checks. They enable the execution of code and the storage of data within the enclaves, protecting them from unauthorised access even by privileged software layers. 
Software components offer a set of APIs and services that enable developers to build secure applications. These APIs ensure that sensitive operations are executed within the enclave. Software components also manage memory isolation, handle enclave initialisation and termination, and enforce access control policies.

Secure boot processes are implemented to ensure that the TEE is launched securely. This involves verifying the integrity and authenticity of the software components involved in establishing the TEE. Secure communication channels, using encryption and authentication protocols, are established to enable secure communication between the TEE and other components of the system, preventing unauthorised access or tampering of data.  Additionally, TEEs often employ attestation mechanisms to verify the integrity of the TEE to external entities. Attestation provides evidence that the TEE is running a genuine and unmodified trusted runtime environment, assuring other parties that they can trust the enclave and its execution environment.

By combining these technical measures, TEEs provide a secure and isolated environment for executing sensitive operations, protecting critical data, and preventing unauthorised access or tampering. 

\subsection{Tickless OS kernels}

When a CPU core is idle or only has a single runnable process, it is unnecessary to generate period timer interrupts because no other task can be switched to. The Linux kernel on x86 from version~2.6.21 can be configured to drop the timer tick to 1~HZ for idle CPUs using \code{CONFIG\_NO\_HZ}. From version~3.10, this has been extended to non-idle CPU cores with \code{CONFIG\_NO\_HZ\_FULL}. The 1~HZ tick is necessary to keep the scheduler alive. Since version 4.17, the residential 1~HZ tick can be offloaded to a global work queue and handled by a housekeeping CPU remotely. The XNU kernel in Mac OS X~10.4, FreeBSD~9 and the NT kernel from Windows 8 onwards all support tickless operations. 

\subsection{Side channel attacks}

Many resources and channels have been exploited to attack TEE enclaves in recently proposed SCAs, \eg the L1 cache, page tables, and branch prediction. Depending on the nature of the exploited side-channel, the SCA is either mounted on the same CPU core executing the victim enclave or from a different core. Moreover, the per-core resources could be exploited with either Multi-Threading (MT) or via interrupts. 

\mypar{Cross-core attacks} Cross-core SCAs exploit the resources shared among all CPU cores, such as the last-level cache~(LLC) and globally shared buffer. The attack given in~\cite{SchwarzWGMM20}, ARMageddon~\cite{ARMageddon}, and XLATE~\cite{SchaikGBR18} attack LLC.
CROSSTALK~\cite{crosstalk} exploits the globally shared buffer, referred to as \emph{staging buffer}, with the CPUID instruction. 
Compared with same-core SCAs, cross-core SCAs tend to be noisy and often have a low leakage rate because the exploited resources are shared among all the running programs. Triggering interrupts is unnecessary in cross-core attacks. However, they need to synchronise the attack and the victim or slow down the victim execution so as to reduce the noise. For example, XLATE creates a bidirectional channel between the attacker and the victim for the synchronisation, and CROSSTALK injects exceptions, \eg page faults, into the victim to slow down its execution.

\mypar{MT-based attacks} Some TEE-enabled CPUs also support multi-threading, e.g., Intel Skylake and AMD EPYC cpu families. When two threads running in the same physical core share per-core resources, a variety of SCAs become possible. For example, the TLB, the L1 cache, the \emph{line fill buffer}~(LFB), and the store buffer are shared between the two logical cores of the same physical core, making MT-based SCAs that exploit these shared resources possible~\cite{WangCPZWBTG17, BrasserMDKCS17, GotzfriedESM17, memjam, EvtyushkinRAP18, Bluethunder, cacheout, ridl, chen2018}. The per-core resources generally have small capacity, and the stored secret data can be easily evicted by other victim instructions. The spy thread must probe the resource with a high frequency and at the right time. To address this challenge, many MT-based SCAs choose to precisely synchronise the spy and the victim thread and slow down the execution of the victim. Most of the MT-based SCAs~\cite{memjam, EvtyushkinRAP18,chen2018} require public APIs from the victim to trigger the target code. To slow down the victim, Bluethunder~\cite{Bluethunder} interrupts the victim with SGX-step, and BranchScope~\cite{EvtyushkinRAP18} suggests to control the scheduling at fine-grain.


\mypar{Interrupt-based attacks} Per-core resources can alternatively be inspected with interrupt-based attacks. Since TEE enclaves rely on the underlying OS or hypervisor to handle interrupts and exceptions, the untrusted OS or hypervisor can preempt enclave execution and exploit its resource usage. 
Compared with the cross-core and MT-based SCAs, interrupt-based SCAs are easier to perform and incur less noise. For example, in controlled-channel attacks~\cite{xu2015controlled, WangCPZWBTG17, BulckWKPS17, LiZLS19, Morbitzer0H19, WernerMAPM19}, page faults and the A/D bit flip are triggered deterministically by the hardware, and the adversary learns the page access patterns of the victim without noise. For other channels, such as the L1 cache, TLB and LFB, the longer the victim enclave runs without interruption, the higher the number of accesses made to the cache, implying higher noise and less temporal resolution. Frequently interrupting the enclave can result in fine-grained side-channel observations. In particular, SGX-Step~\cite{vanbulck2017sgxstep} enables the adversary to configure the advanced programmable interrupt controller~(APIC), thus interrupting SGX enclaves at instruction granularity. A range of other attacks on Intel SGX~\cite{ZombieLoad, LVI, foreshadow, BulckPS18, frontal, Bluethunder, CopyCat} are built on SGX-Step, interrupting the enclave after each CPU cycle. Load-step~\cite{Loadstep} and SEV-step~\cite{wilke2023sevstep} are similar tools that can single-step ARM TrustZone and ARM SEV enclaves. 

Due to their lower noise and more fine-grained control, interrupt-based attacks are more powerful and achieve a higher leakage rate. 
It is crucial to have an effective countermeasure to defend against such powerful interrupt-based attacks.

\subsection{Existing defences}

At a high level, existing mitigations against SCAs either work at the hardware or software level. The techniques used in software solutions include randomization, normalisation, and interrupts detection. 

\myparr{Hardware-based} defences propose new hardware designs to eliminate certain types of SCAs. To mitigate page-based SCAs for Intel SGX, Autarky~\cite{OrenbachBS20} proposes to modify the SGX architecture so that the enclave no longer reports the faulting page base address to the untrusted OS. To prevent single-stepping enclaves with interrupts, AEX-Notify~\cite{AEX-Notify} proposes a hardware ISA extension for Intel SGX. 
Hardware-based defences incur lower performance overhead compared to most software-based defences, but they cannot be applied to existing platforms.


\myparr{Normalization and randomization} generalise the software-based hardening techniques that eliminate secret-dependent control flow and make data access \emph{oblivious}. Normalization makes sensitive portions of the enclave code behave identically for all possible inputs. 
In contrast, randomization approaches randomize the code and data access. ORAM and \emph{address space layout randomization}~(ASLR)~\cite{NurmukhametovZK18,ShachamPPGMB04,GiuffridaKT12} are two traditional techniques to randomize code and data.
The advantage of these approaches is they do not have false negatives or positives. However, they heavily rely on code transformation and impose high overhead on both the performance and space. 

\myparr{Detection} injects code into the enclave to detect SCAs and terminate the enclave execution if attacked. As mentioned the same-core SCAs are either MT-based or interrupt-based, and existing detection techniques either detect one or the other. 
To mitigate MT-based SCAs, Varys and Hyperspace run a shadow thread with each enclave thread on the same physical core, and check if the threads indeed share a core. To detect interrupt-based SCAs for Intel SGX, \emph{transactional synchronization extension}~(TSX)~\cite{HerlihyM93} is widely used because interrupts or conflicts are illegal within a TSX transaction. T-SGX~\cite{Shih0KP17} uses TSX to detect page faults.
Varys~\cite{varys} periodically detects if an interrupt has occurred and aborts the enclave if interrupts occur too frequently. Amongst the existing defences, only Varys can completely prevent all interrupt-based SCAs.

\mypar{Discussion} Compared with normalization and randomization techniques, detection solutions have a lower performance overhead. However, they still need to inject detection code into the enclave, which causes performance and space challenges. Moreover, they are prone to miss malicious interrupts or misclassify regular interrupts as attacks, \ie suffer from false negatives and positives. Oleksenko \etal{}~\cite{varys} evaluate Varys by detecting interrupts every 3400~instructions, which causes a 15\% performance overhead. Indeed, SGX-step can interrupt the enclave after each instruction, which means that 3399~interrupts can escape each detection cycle. To reduce false negatives, Varys can detect interrupts more frequently, yet with heavier overheads.


Interrupt-based attacks are powerful, yet existing defences suffer from one or more of the following limitations: \ding{182}~high false negatives; \ding{183}~only applicable to specific attacks or hardwares; and \ding{184}~performance and space overheads. 
The goal of our work is to design a defence without the above limitations. We aim to design an efficient pure software-based defence that can protect enclaves from all the interrupt-based attacks with fewer false negatives.


%% file: sections/overview.tex

\section{Design of \sys}
\label{sec:overview}

In this section, we discuss our threat model and assumptions and design details of \sys.

\subsection{Threat model}
Our primary goal is to protect enclaves against a majority of side-channel attacks that require interrupts. 
We consider a privileged adversary who aims to extract sensitive information from an enclave through interrupts-based SCAs. 
We assume that the adversary completely controls the OS or hypervisor and has every capability an OS or hypervisor may have over an application, excluding those restricted by TEE. In particular, the adversary may mount any interrupt-based attack and interrupt the enclave execution precisely, as required, \eg at instruction-level granularity. 

Furthermore, the adversary has access to the enclave source code and/or binary, so that they know the detailed behavior of the target enclave, \eg related to secret-dependent control flow and memory access patterns. However, the adversary cannot directly access the memory of the enclave and any internal CPU state (\eg CPU registers). The adversary also cannot modify the code or initial enclave data, because the enclave's integrity can be verified using remote attestation.



Denial-of-service~(DoS) attacks are beyond the scope of the paper, as a malicious OS can always decide to terminate the enclave.

%% file: sections/details.tex
\begin{figure}
\centering
\includegraphics[width=0.4\textwidth]{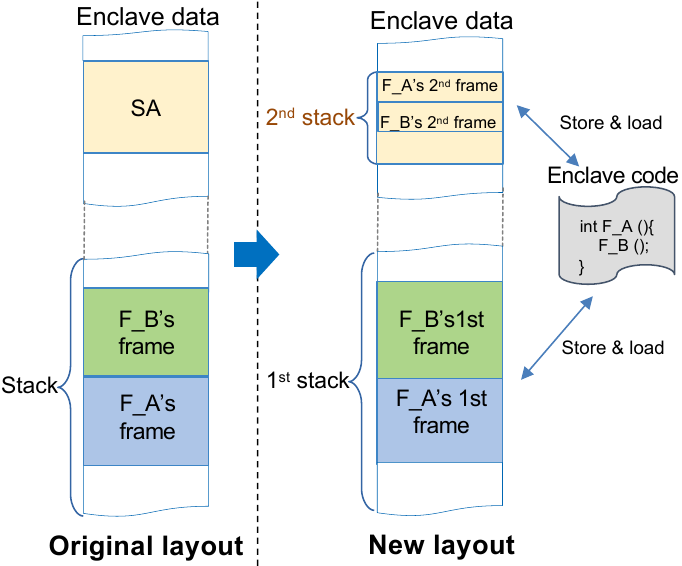}\\
\caption{Data layout of \sys. \textnormal{The SA is used as a second stack.
There is a frame in the SA for each function call to store memory references. The enclave code is instrumented to load/store memory references from/into the SA.}}
\label{Fig:basic}
\end{figure}

\subsection{Interrupt-free environment}
\label{subsec:interrupt}

\sys is designed to detect malicious interrupts, meanwhile it should ensure that no innocent interrupts affect enclave execution. TEEs currently cannot distinguish the origins of interrupts, and it is non-trivial to decide if an interrupt is triggered by an attacker or not. \sys therefore requires an interrupt-free environment to run the enclave successfully when there is no attack.

Interrupts are events that indicate that a condition exists somewhere in the system, the processor, or within the currently executing program that requires the attention of a CPU. They typically result in a forced transfer of execution from the currently running program to an interrupt handler. The interrupts received by CPU could be generated from software or external hardware. Software interrupts are triggered by the CPU itself when executing instructions, \eg system calls, page faults and a divide-by-zero exception, and they typically are called \textit{traps} or \textit{exceptions}. Hardware interrupts are caused by the state of hardware devices external to the CPU. For example, a network interface card~(NIC) generates an interrupt to signal that a packet has arrived. 


\sys requires that the enclave itself does not contain faults, errors, or instructions that could cause exits\footnote{\sys can remove this requirement by dumping the content in the SA to another place before executing such instructions and copying data back afterward. We will work on it in the future work.}. An enclave may trigger exits when processing local timer ticks or other interrupts, \eg from I/O devices. 
The local timer ticks can be disabled by using the latest kernels and putting the CPU core into the tickless mode. Most other types of interrupts can be pinned to other cores. 

The tickless mode cannot disable the interrupts due to TLB shootdowns. When a CPU core changes the virtual-to-physical mapping of an address, it needs to send inter-processor interrupts~(IPIs) to all online cores, including the tickless core, to invalidate that mapping in their TLBs. As TLB shootdowns are expensive in performance,  techniques~\cite{AwadBBSL17, DiDi, KumarMKVYKBK18, Amit17}, such as batching and self-invalidating, have been proposed to reduce or avoid them. These techniques can be applied to \sys to remove TLB shootdowns from tickless cores.

\subsection{The SA canary}
\label{subsec:ssa}

The above techniques make an interrupt-free environment available to \sys, yet a malicious OS may invalidate it easily, \eg by intentionally turning off the tickless mode. The enclave is supposed to be executed in an interrupt-free environment. Any interrupt occurred to the enclave is regarded as an attack. \sys therefore aborts the enclave once it detects an interrupt. As shown in Fig.~\ref{Fig:basic}, \sys achieves that by utilising the SA as a \emph{second stack} to instrument each enclave function to store and load memory references to and from the SA, \eg each function's frame base address. 

The CPU enforces restrictions on memory access, and minor changes to an address may violate such restrictions. For example, the enclave can only access user space memory. The CPU raises an exception and aborts the enclave if this requirement is not satisfied. Such an approach has two challenges: \ding{182} TEE must modify the addresses stored in the SA into illegal or unreachable forms, rather than a form that can instruct the enclave to execute incorrect code or access other memory locations; \ding{183} \sys must timely response to the attack, \ie ensure that the addresses stored in the SA are accessed as soon as possible after resuming, otherwise the enclave will not abort or abort too late. To address the first challenge, \sys ensures that TEE can deterministically or very likely modify the addresses into \emph{non-canonical form} on interrupts.



\mypar{Non-canonical addresses} \sys is designed specifically for 64-bit architectures, in which virtual addresses must be of a \emph{canonical form}. Specifically, x86-64 processors with 4-level paging (known as LA48) require bits 48 through 63 of a virtual address to be copies of bit 47, \ie the most significant 16-bit of a virtual address must be \code{0x0000} or \code{0xffff}; x86-64 processors with 5-level paging (known as LA57) require bits 57 through 63 to be copies of bit 56. In the following, we assume the CPU uses $v$~bits for addressing, and the most significant $u$~bits must be copies of bit $v-1$, where $v+u=64$. Accessing a non-canonical address aborts the enclave. \sys stores addresses in a way that they will be changed to non-canonical form by TEE and relies on the \emph{extended features} supported by the CPU.  

\mypar{CPU extended features} To speed up the operations performed on multiple data objects, modern CPUs generally support several sets of extensions to x86 instruction set architecture, including Single Instruction Multiple Data (SIMD), Intel Streaming SIMD Extensions (\eg \code{SSE}, \code{SSE2}, \code{SSE3}, and \code{SSE4}), and Intel Advanced Vector Extensions (\eg \code{AVX}, \code{AVX2}, and \code{AVX-512}). The CPUs with TEE support all or multiple of them. For instance, Intel Skylake, Comet Lake, and AMD EPYC CPUs support all the versions of \code{SSE}, \code{AVX} and \code{AVX2}. ARM CPUs with TrustZone technology also include the NEON SIMD feature. Additional registers, \eg \code{XMM0-XMM15}, \code{YMM0-YMM15}, and \code{ZMM0-ZMM31}, or \code{Q0-Q31}, are added into the CPU to support these features, and the SA has regions to save them when the enclave exits. Data-intensive applications, such as 3D graphics and video processing, use these features for faster processing.

\mypar{Deterministic address modification} \sys is designed based on the observation that not all the extended features supported by the CPU are used by a single program, \ie a set of registers are never used during the execution of a program. Our idea is to identify the extended features supported by the CPU but not used by the enclave and utilise their corresponding regions in the SA to store memory references. For instance, assume \code{AVX-512} is the identified feature, within the SA the region used to store \code{ZMM16-ZMM31} register state will store addresses in \sys. Moreover, in order to ensure the addresses stored there will be modified into non-canonical form, \sys primes every 64-bit of the extended feature registers with a non-canonical address, \eg \code{0xAAAAAAAAAAAAAAAA}. Considering these registers will never be used by the program, the values stored inside them will never be modified in any case. When the enclave is interrupted, the addresses stored in the second stack will all be overwritten with \code{0xAAAAAAAAAAAAAAAA}, which is non-canonical. 

\mypar{Exception} In case all the CPU registers are used by an application, \sys ensures the enclave terminates upon interrupts with a high probability. When the enclave is interrupted, the stored CPU state is not deterministic. Only if a register stores an address (we call it address-state), its most significant $u$~bits must be all $0$ (enclave runs in userspace). Otherwise, any continuous $u$~bits may be all $0$ with a probability of $1/2^{u}$, which is $1/2^{16}$ and $1/2^{7}$ for LA48 and LA57 architectures, respectively\footnote{Note that due to the remote attestation, attackers cannot modify the enclave code and data to enforce non-address-states to have continuous $u$~bits $0$s or $1$s.}. When addresses stored in the SA are overwritten by address-state in an unaligned way, they will be modified to illegal form with a probability of $1-1/2^{u}$, \ie $99.998\%$ and $99.2\%$ for LA48 and LA57, respectively. 
As such, we store data into the SA with an $o$-bit offset, where $u \leq o \leq v$. By doing so, the most significant $u$-bit of addresses stored in the SA is overwritten by the bits between bit $63 - o - u$ and $63-o$ of registers, which are all $1$ or $0$ bits with a probability of $1/2^{u}$, no matter if they are address-state or not. 


\begin{figure}
\centering
\includegraphics[width=0.47\textwidth]{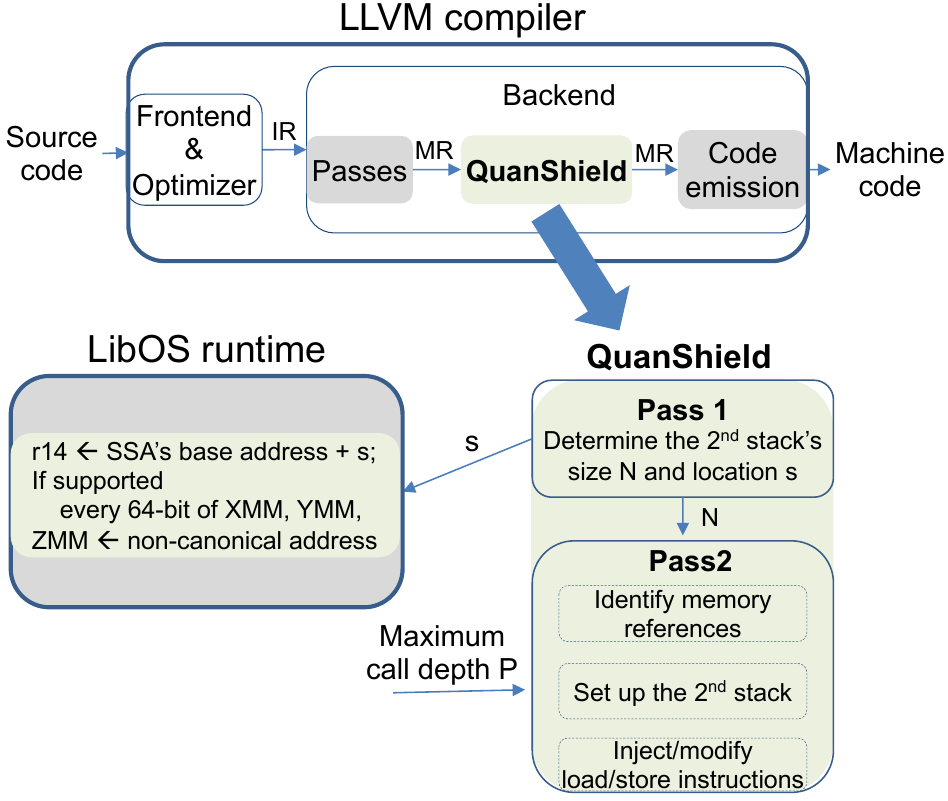}\\
\caption{The components of \sys. \textnormal{\sys adds 2 new machine function passes to the LLVM compilation framework, which operate on LLVM's machine representation (MR) code. \sys reserves \code{r14} and uses it as the frame base pointer of the 2nd stack.}}
\label{fig:components}
\end{figure}
\begin{figure}
\centering
\begin{lstlisting}[language={[x64]Assembler}, frame=single]
#%bb.0:
	pushq %rbp 
	.cfi_def_cfa_offset 16
	.cfi_offset %rbp, -16
	movq  %rsp, %rbp
	.cfi_def_cfa_register %rbp
	subq. $32, %rsp
	movq  %rdx, -24(%rbp)
	...
	
#%bb.2:
	movq   -8(%rbp), %r11
	...
	movq   -24(%rbp), %rax
	movslq -32(%rbp), %rcx
	cmpq   $1, (%rax, %rcx, 4)
	jne   .LBB2_4
	...
	
.LBB2_6:
	...
	popq %rbp
	retq
\end{lstlisting}
\caption{Original code example}
\label{code:original}
\end{figure}
\mypar{Enclave termination} After resuming, the enclave proceeds until accessing a non-canonical address. \sys could just rely on the enclave itself instructions to ensure termination (only the source or destination address is changed), which hardly affects the enclave's performance since no additional instruction is injected. For instance, in the original code, each function's caller's frame base address is stored into the stack at the beginning (\eg Line 2 in Fig.~\ref{code:original}), 
loaded into \code{rbp} at the end of the function (\eg Line 22 in Fig.~\ref{code:original}), and accessed by its caller. By storing/loading each caller's frame base address into/from the SA, rather than the original stack, if the callee is attacked, the enclave will abort when the caller accesses its stack (because the caller's frame base address reloaded into \code{rbp} from the SA has been modified into non-canonical from by the hardware). However, in this case, if the caller does not access the stack after the function call or accesses the stack too late, the attacker is still able to get useful information. To address the second challenge, \sys can achieve stronger protection with a bit higher overhead by partially relying on the enclave itself instructions and injecting additional instructions \footnote{All existing detection-based solutions rely on instruction injection.}. Here we provide block-granularity protection as a strawman example. 

\mypar{Block-granularity protection} By investigating the code of several popular attack targets, such as the cryptographic algorithms in OpenSSL~\cite{OpenSSL}, we observe that the entropy of a single block is limited and the secret information is usually related to multiple blocks. So \sys basically provides \emph{block-granularity} protection: no matter which code block is interrupted, the enclave is guaranteed to terminate before executing its next block. \sys achieves that by injecting one or two instructions per block. Precisely, \sys instruments each function to store its own frame base address to the SA and reload it to \code{rbp} from the SA before executing each code block. For the blocks whose first instruction is not frame access, \sys injects a dummy one. If the enclave has been attacked, the reloaded address must be in non-canonical form and lead to termination when accessing it. Since the non-canonical address is independent of blocks, the attacker cannot learn which block is executed after the attacked one, indicating the control flow of inter-blocks is also protected. The protection granularity can be improved further by reloading the frame base address from the SA more frequently, \eg before each stack access, but incurring higher overhead. 

\mypar{Intra-block protection} Frame base address is usually held by a dedicated register (\eg \code{rbp}), and it is rarely to be stored and loaded (only when a function is called and completed). In \sys, to reflect the change of the SA, the enclave is instrumented to load it frequently from the SA, which adds performance overhead. Indeed, the enclave also has many other memory references that cannot held by dedicated registers and have to be loaded into a temporary register before accessing it in the original code (\eg Line 14 in Fig.~\ref{code:original}). When the second stack has sufficient capacity, \sys can store and load those generic memory references from the SA, which could achieve intra-block protection without overhead since no additional instruction is required. 

%
                      

%% file: sections/implementation.tex
\section{Implementation for Intel SGX}
\label{sec:implementation}
In this section, we take Intel SGX as an example to show the specific implementation details of \sys. 

\subsection{Intel SGX}
\begin{figure}
\centering
\includegraphics[width=0.1\textwidth]{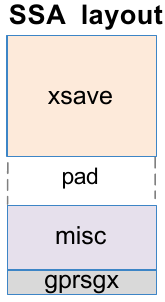}\\
\caption{The SSA layout of Intel SGX. \textnormal{\code{XSAVE}  area typically ranges between 2~KB and 3~KB, and \code{MISC} ranges between 512 bytes and 1~KB. \code{GRPSGX} area is 176 bytes.}}
\label{fig:ssa}
\end{figure}



Intel Software Guard Extensions~(SGX) is Intel's TEE platform. 
SGX enclaves rely on the underlying OS to handle interrupts and faults. During enclave execution, interrupts or exceptions cause the enclave to perform an \emph{asynchronous enclave exit}~(AEX) which saves the CPU state in a \emph{state save area}~(SSA) inside enclave memory before giving CPU control to the OS. 

As shown in Fig.~\ref{fig:ssa}, each SSA frame consists of \code{XSAVE}, \code{MISC}, \code{GRPSGX}, and \code{pad} regions. \code{XSAVE} and \code{MISC} store the registers used for extended features supported by the CPU, such as \code{FP0-FP7} and \code{XMM0-XMM15}. Their sizes are determined by the features supported by the CPU and can be reduced by the enclave implementor. 
\code{GRPSGX} is used to store the general-purpose registers, \eg \code{rip}, \code{rsp} and \code{rbp}. The size of an SSA frame must be in pages and large enough to hold all chosen states. The \code{pad} region is only used to ensure that the SSA is aligned to the end of a 4~KB page. 
An interrupted enclave can be continued by means of the \code{ERESUME} instruction. \code{ERESUME} restores the state from the SSA into the CPU after the interrupt or fault has been handled.


\subsection{Library OS runtime systems}

SGX enclaves cannot directly access hardware such as network or storage devices. 
They are required to cooperate with the untrusted host to perform system calls or access hardware directly. Enclave applications can be developed using the Intel SGX SDK~\cite{SGXSDK} or the Open Enclave SDK~\cite{OpenEnclave}, which take these constraints into account. 
Library OS runtime systems for SGX, \eg Haven~\cite{baumann2014haven}, Graphene-SGX~\cite{tsai2017graphene}, SCONE~\cite{arnautov2016scone} and SGX-LKL~\cite{SGXLKL}, remove such constraints by running a library OS inside the enclave. They thus allow legacy applications to execute inside enclaves without code modifications. Moreover, some of them, \eg SGX-LKL and SCONE, can avoid context switches for system calls. 


SGX-LKL uses a port of the Linux kernel inside an enclave as a library OS to provide full POSIX compatibility. It provides most of the OS functions inside the enclave and also includes a libc implementation to support unmodified binaries that are dynamically linked against libc. SGX-LKL only requires 7~calls in its host interface. To avoid costly context switches between the enclave and the host when calling the host interface, SGX-LKL creates host threads outside of the enclave, which handles host calls asynchronously without triggering AEXs. The threads inside of the enclave always execute application tasks. 
\sys relies on SGX-LKL to execute programs within Intel SGX.


\subsection{Implementation details}
%
%
%
\begin{figure}
\centering
\begin{lstlisting}[language={[x64]Assembler}, frame=single, escapechar=!]
#%bb.0:
!\colorbox{yellow}{addq~  \$16, ~\%r14}!;Setup the 2nd frame, r14 is the 2nd frame's base pointer
	pushq %rbp
	.cfi_def_cfa_offset 16
	.cfi_offset %rbp, -16
	movq  %rsp, %rbp
!\colorbox{yellow}{movq~ \%rbp, ~(\%r14)}!;Store frame base address to SSA 
	.cfi_def_cfa_register %rbp
	subq $32, %rsp
!\colorbox{orange}{movq~ \%rdx, ~-8(\%r14)}!;Store generic memory reference to SSA	
	...
    
#%bb.2:
!\colorbox{yellow}{movq~ (\%r14), ~\%rbp}!;Load frame base address from SSA
	movq   -8(%rbp), %r11;Could crash here
	...
!\colorbox{orange}{movq~ -8(\%r14),~ \%rax}!;Load generic memory reference from SSA
	movslq -32(%rbp), %rcx;
	cmpq $1, (%rax, %rcx, 4);Could crash here
	jne .LBB2_4
	...
	
.LBB2_6:
!\colorbox{yellow}{movq~ (\%r14), ~\%rbp}!;Load frame base address from SSA 
!\colorbox{yellow}{movq~ -8(\%rbp), \%r10}!;Dummy access, could crash here
	...
	popq %rbp;
!\colorbox{yellow}{subq~ \$16, ~\%r14}!;Setup the 2nd frame
	retq
\end{lstlisting}
\caption{\sys instrumentation with block-granularity protection. \textnormal{The instructions highlighted in yellow are \textbf{injected} by \sys, and those highlighted in orange are \textbf{modified} by \sys.}} 
\label{code: implementation}
\end{figure}
We create a prototype for \sys based on the LLVM compiler, which can handle arbitrary C and C++ code. As shown in Fig.~\ref{fig:components}, we extend the LLVM compilation framework by adding two new machine function passes to the LLVM backend, which operate on LLVM's machine representation~(MR) code. 

The first pass is used to determine the size and location of the 2nd stack in the SSA. It first gets the extended features supported by the CPU with \code{CPUID} instruction and then identifies the registers not used by the program. Based on that, it identifies the largest continuous region in the SSA that will be touched by AEXs but not affected by the program. This region will be used as the 2nd stack, and its offset to the SSA base address, \code{s}, is sent to SGX-LKL. If all the extended features are used by the program, the 2nd stack will start from the largest continuous region's base address with $o$-bit offset. The size of the 2nd stack, $N$-byte, is sent to the second pass.

The main functionality of the second pass is to instrument each function to store/load memory references into/from its second frame. It basically processes each function in 3 phases: (i)~identify memory references; (ii)~set up the second frame; (iii)~inject and modify load and store instructions. For clarity, we take the code shown in Fig.~\ref{code:original} as an example to illustrate the implementation details. The code instrumented by \sys with block-granularity protection is shown in Fig.~\ref{code: implementation}.

\mypar{Memory reference identification} Other than frame base address, we also store generic memory addresses to the SSA since this operation can improve security level without additional overhead. The purpose of this phase is to identify how many memory references each function used and where they are stored. Based on the fact that memory references stored in the stack must be loaded into registers for accessing data there, our main idea is to check if any register (other than \code{rbp} and \code{rsp}) is used as the \code{BaseReg} of a memory operand. If yes, the register must hold a memory reference. For each such register, the pass then tracks where the memory reference is loaded from and records the source location. For instance, in the example code shown in Fig.~\ref{code:original}, only \code{rax} is used as \code{BaseReg} in line 16, and it is loaded from \code{-24(\%rbp)} in line 14, thus the location \code{-24(\%rbp)} will be recorded for later use. The number of memory references will be used to set up the 2nd frame's size, and the recorded source locations will be modified in the last phase.  
    
%
 
\mypar{Second frame setup} \sys requires 2 instructions to set up the second frame, \ie the instructions in lines~2 and~28 in Fig.~\ref{code: implementation}, which involve two operands: the second frame's base pointer and size.

\sys reserves one general-purpose register, \eg \code{r14}, as the second frame's base pointer, and it is initialized with the base address of the 2nd stack within SGX-LKL before running the program. SGX-LKL also initializes every 64-bit of the supported extended features' registers with a non-canonical address.

Each function's second frame size varies with the number of memory references it has and is limited by the total size of the 2nd stack and the maximum function call depth of the program. Assume they are $M$, $N$-byte, and $P$, respectively. To avoid stack overflow, \sys limits each frame on the 2nd stack to $\frac{N}{P}$-byte. Thus, each function's 2nd frame size will be $8*\min\{(M+1), \lfloor\frac{N}{8P}\rfloor\}$-byte. If $M+1 > \lfloor\frac{N}{8P}\rfloor$, only $\lfloor \frac{N}{8P}\rfloor-1 $ generic addresses can be stored in the SSA. Currently \sys relies on the program developer to provide the value of $P$.

\mypar{Instruction injection and modification} This phase aims to inject the instructions to store and load frame base address into/from the SSA and modify the source/destination addresses of the instructions that load/store generic memory references. 

During the original frame setup, \code{rbp} will be assigned the function's frame base address. The pass first identifies that instruction and injects an instruction to store \code{rbp} into the 2nd frame after it, \eg line 7 in Fig.~\ref{code: implementation}. Second, for each block, the pass injects an instruction to reload the frame base address into \code{rbp} at the beginning, \eg line 14 and 24 in Fig.~\ref{code: implementation}. Moreover, the pass checks if each block's first instruction is stack access. If not, \sys injects a dummy one, \eg line 25. By doing so, each block will only proceed when its previous blocks are not attacked. This part injects 1 or 2 instructions into each block, which is the main overhead added by \sys.

For generic memory references, \sys just needs to modify existing instructions. The first phase has already figured out the locations that store memory references. This phase randomly picks $\min\{M, \lfloor \frac{N}{8P}\rfloor-1\}$ of them. For all instructions involving those locations as operands, the pass replaces them with locations in the 2nd frame. For instance, in Fig.~\ref{code:original}, line 8 and 14 involves \code{-24(\%rbp)}, and they are replaced with \code{-8(\%r14)} in Fig.~\ref{code: implementation}.


%% file: sections/evaluation.tex

\section{Evaluation}
\label{sec:evaluation}

We evaluate the performance, false negatives, and false positives of \sys with several different applications.

\subsection{Experiment setup}

The experiments mainly run on a machine with the 3rd generation Intel Xeon scalable processors silver 4310. The EPC size is 8~GB. The machine uses Ubuntu Linux~18.04 and SGX driver~v2.5. The Linux kernel~v4.19 is set up to provide an interrupt-free environment. We turn on the tickless mode and turn off the hyperthreading feature. To disable shootdowns on tickless cores, we modify the kernel to exclude tickless cores from executing \code{flush\_tlb\_func\_remote} function. The LLVM framework is~v9.0.1. 


\subsection{Performance overhead}
%
%
\begin{table}[]
\centering
\scriptsize
\caption{Throughput of Nbench.}
\label{tbl:nbench-perf}
\begin{tabular}{l|c|c|cccc}
\toprule
\multirow{3}{*}{\textbf{nbench}}
& \multirow{3}{*}{\textbf{\#Blocks}}
&{\multirow{3}{*}{\begin{tabular}[c]{@{}c@{}} \textbf{\sys} \\ \textbf{slowdown} ($\%$) \end{tabular}}}                                              
& \multicolumn{4}{c}{\multirow{2}{*}{\textbf{Varys slowdown ($\%$)}}}                           
\\
&
&
\\
                          
&  
& 
& \textbf{I=4}
& \textbf{I=8}
& \textbf{I=16}
& \textbf{I=32}

\\ \midrule
Fourier                 
&86
&1.65
&4.94 &1.81 &1.45 &1.14  
\\
Bitfield
&88
&12.5
&32.1 &22 &7.9 &4.47
\\
NumSort 
&90 
&12.1
&35.4 &19.3  &18.2 &11.4                              
\\
Idea 
&130
&1.73
&63.4 &39.6 &28.6 &19.2
\\
Huffman                 
&159
&10.1
&53.1 &37.1 &14.7 &9 
\\
StringSort             
&179
&16
&17.2 & 12.9 &5.03  &2.52 
\\
NNet
&195
&16.3
&51.1 &36.9 &17.9 &18 
\\
LU   
&200
&16.1 
& 57.5 &45.7 &30.4 & 22.5
\\
Assignment              
&296
&11.1
&52 &32.4 &26.5 &16 
\\
FP emulation 
&480           
&11.6
&44.5 &23.9 &15.7 &1.2                
\\\hline
Average
&190.3
&10.3
&43.7 &28.7 &17.9 &11.4
\\ \bottomrule                 
\end{tabular}
\end{table}
\begin{table}[]
\centering
\scriptsize
\caption{Runtime of PARSEC.}
\label{tbl:parsec-perf}
\begin{tabular}{l|c|c|cccc}
\toprule
\multirow{3}{*}{\textbf{Parsec}}
& \multirow{3}{*}{\textbf{\#Blocks}}
&{\multirow{3}{*}{\begin{tabular}[c]{@{}c@{}} \textbf{\sys} \\ \textbf{slowdown} ($\%$) \end{tabular}}}                                              
& \multicolumn{4}{c}{\multirow{2}{*}{\textbf{Varys slowdown ($\%$)}}}                                        
\\
&
&
\\
                          
&  
& 
& \textbf{I=4}
& \textbf{I=8}
& \textbf{I=16}
& \textbf{I=32}

\\ \midrule
Blacksholes
&19
&3.93
&31.4 &20.9 &9.49 & 4.67
\\
Swaptions
&215
&23.8
&106 &65.2 &31.5 &13.1
\\
Dedup 
&532
&0.32
&32.1 & 17.4 & 6.42 &2.14
\\
Fluidanimate
&618
&7.68 
&61.6 &33.5 &20.7 &3.2
\\
Freqmine
&1215
&30.3
&67.1 &39.8 &18.3 &9.69
\\
Canneal
&1588
&12.5
&60.2 &19.6 &5.53 &1.48 
\\
Bodytrack
&5233
&23.7
&129 &50.1 & 19.2 & 7.78
\\
X264
&14481
&14.8
&217 &98.2 &46.6 &7.4
\\ \hline
Average
&2987
&14.6
&87.2 &37.8 &25.6 &6.61
\\ \bottomrule                 
\end{tabular}
\end{table}
%

                          


The main overhead of \sys is due to the injected instructions. 
\sys inserts 2 instructions per function for setting up the second stack, inserts 1 or 2 instructions per code block to reload frame base address from the SSA. 

As done in many previous works~\cite{FuBQL17, Oleksii, Shih0KP17, ChenZRZ17, Hyperspace, drsgx, OrenbachBS20}, we evaluate the performance of \sys with Nbench~\cite{nbench} and Parsec~\cite{parsec} benchmarks, and compare the runtime of \sys enclaves with a baseline and Varys. The baseline runs the original application without any protection. The source code of Varys is not publicly available. We implement a prototype of Varys based on the description in the paper. Specifically, we initialise a field of SSA with a specific value and insert the code to check if this field is modified per $I$~IR instructions, where $I$ is a configurable number, and the enclave exits as long as an AEX is detected\footnote{Cache eviction with dummy data and the handshake for co-location test are not implemented in our prototype as they are orthogonal to \sys. No threshold is required since Varys is also tested in an interrupt-free environment.}. In our tests, $I$ varies from 4 to 32. All the test cases are executed in an interrupt-free environment with SGX-LKL. The runtimes reported below are averaged over 10 runs.

%
%

The test results of Nbench and Parsec are shown in Table~\ref{tbl:nbench-perf} and ~\ref{tbl:parsec-perf}, respectively. The results show that the performance of Varys changes linearly with the value of $I$. The performance of \sys roughly increases with the number of blocks, yet it also depends on how many dummy instructions are injected and how many blocks are executed during the test. In most cases, \sys outperforms Varys with $I=16$. On average, when the task has around 190 and 2987 blocks, \sys adds 10.3\% and 14.6\% overhead which is better than Varys with $I=32$ and $I=16$, respectively. 

\subsection{Security evaluation}
For the security guarantee of \sys, we first evaluate the rate of enclave crash if it is attacked even only with one interrupt. The results show that all the tested workloads 100\% crash when they are protected with \sys. 

We also measure \sys{}'s response delay to attacks by counting the number of interrupts the attacker can trigger before an enclave crash. 


\mypar{Attack setup} In the experiment, with SGX-Step, we configure the local APIC of the tickless core and periodically deliver interrupts to the enclave running on it at instruction granularity. In addition, we register an interrupt handler in userspace to count AEXs. In our test, the attacker greedily interrupts the enclave as frequently as possible. By tuning the APIC timer interval and monitoring \code{rip} upon each AEX repeatedly, we find that 45 is the proper APIC timer interval in our testbed that makes the enclave progress under the attack (\ie avoid zero-step). 

We test the response delay of \sys in two cases: extended features are not used and used by the enclave. The first case is tested with Nbench and OpenSSL~\cite{OpenSSL}. The second case is tested with OneDNN~\cite{onednn}, a library with basic building blocks for deep learning applications, and Eigen~\cite{eigen}, a library for linear algebra. For OneDNN and Eigen, we test the functions using the SSE or AVX features and locate the second stack at \code{GRPSGX} region with an offset of 16-bit. 

We run each test case under attacks 100~times and each time the attack starts at random places. Each time we record the number of AEXs occurred before the enclave terminates. The reported values are the average over 100 runs.

\begin{table}[]
\centering
\scriptsize
\caption{Attack response delay for Nbench.}
\label{tbl:nbench-sec}
\begin{tabular}{l|c|c|cccc}
\toprule
\multirow{2}{*}{\textbf{Nbench}} &
\multirow{1}{*}{\textbf{Block}} &
\multirow{2}{*}{\textbf{\sys}} 
& \multicolumn{4}{c}{\textbf{Varys}} 
\\
&\textbf{Size}
&
& I=4    & I=8     & I=16  &I=32     
\\ \midrule
Assignment  
&2.6
&4.64
&6.67 & 14.7 &18.4 &27.9
   \\
LU 
&4.3
&4.91
&6.92 &14.8 &17.5 &26.6
\\
NNet 
&5.3
&6.59
&6.56 & 13.6 & 17.8 &33.5
\\
Bitfield  
&6.1
&5.99
&5.92 &11.8 & 16.6 &31.6
 \\
NumSort 
&6.2                         
&6.16 
&5.77 & 11.2 & 17.4 &33.9
\\
StringSort  
&8.5
&6.85
&5.76 & 10.9 & 17.2 &34.5
  \\

Huffman
&12.6
&11.4
&6.14 &13.4 &16.8 &23.2
 \\
Idea 
& 40.6
& 14.6
& 6.12 & 13.7 &19.9 &35.5
\\ \bottomrule     
\end{tabular}
\end{table}

\begin{table}[]
\centering
\scriptsize
\caption{Attack response delay for OpenSSL.}
\label{tbl:OpenSSL}
\begin{tabular}{l|c|c|cccc}
\toprule
\multirow{2}{*}{\textbf{Parsec}} &
\multirow{1}{*}{\textbf{Block}} &
\multirow{2}{*}{\textbf{QuanShield}} 
& \multicolumn{4}{c}{\textbf{Varys}} 
\\
&\textbf{size}
&
& I=4    & I=8     & I=16  &I=32     
\\ \midrule
ECDSA  
&3.3
&6.85
&5.93 &11.9 & 17.8 &33.1
 \\
DH  
&3.6
&6.79
&5.38 &10.7 & 15.5 &30.7
\\
RSA  
&3.7
&5.94
&5.65 &12.9 &19.7 &35.2
\\
DSA  
&3.8
&6.13
&5.14 &10.9 &18.4 &31.3
\\
HMAC-SHA256  
&3.8
&6.1
&5.89 &13.4 &18.8 &33.9
\\
HMAC-MD5
&3.9
&6.54
&5.76 &13.7 &18.1 &33.8
\\
AEX Enc 
&5.1
&7.64
&5.93 &10.4 &17.6 &32.9

\\ \bottomrule     
\end{tabular}
\end{table}
\begin{table}
\centering
\scriptsize
\caption{Attack response delay for OneDNN and Eigen.}
\label{tbl:onednn}
\begin{tabular}{l|c|c|cccc}
\toprule
\multirow{2}{*}{\textbf{}} &
\multirow{1}{*}{\textbf{Block}} &
\multirow{2}{*}{\textbf{QuanShield}} 
& \multicolumn{4}{c}{\textbf{Varys}} 
\\
&\textbf{size}
&
& I=4    & I=8     & I=16  &I=32     
\\ \midrule
RNN inference  
&9.6
&16.5
&6.09 &12.1 &17.8 &31.5
 \\

CNN training  
&16.3
&20.9
&5.88  &12.3 &18.5 &34.9
 \\

CNN inference  
&17.5
&21.8
&5.12 &11.6 &17.3 &32.3
\\

JacobiSVD
&8.9
&6.84
&6.82 & 8.84 &25 &31.5
\\

BDCSVD
&6.74
&2.52
&3.68 &7.4 &9.4 &29.9
\\ \bottomrule     
\end{tabular}
\end{table}
%



%
\mypar{Results}The results are shown in Table~\ref{tbl:nbench-sec}, ~\ref{tbl:OpenSSL} and~\ref{tbl:onednn}. In most cases, the response delay of \sys roughly matches the block size and is comparable to Varys when $I$ closes to the block size. With a comparable security guarantee, \sys adds less overhead than Varys, because \sys injects at most 2 instructions per block, yet Varys injects at least 3 to check the SSA for every $I$ instructions. For instance, when the block contains less than 9 instructions on average, \sys will terminate the enclave within 7 interrupts, which is comparable to Varys with $I=4$. Tables~\ref{tbl:nbench-perf} and~\ref{tbl:parsec-perf} show that \sys outperforms Varys with $I=4$ by up to $2.8 \times$. 

Since \sys also stores generic memory references in the SSA, when the attacked functions have enough generic memory references, the response delay of \sys will be much shorter. For instance, the tested function of \textit{Idea} contains 2 generic memory references stored in the SSA, and it can crash within 15 interrupts although its average block size is $40.6$. 










%

\mypar{False positives} We also evaluate the false positives of \sys for PARSC and Nbench by running each workload 1000 times in an interrupt-free environment and measuring the failures. The result shows that the false positive rate for all the tests is 0.


%% file: sections/relatedwork.tex

\section{Related work}

\subsection{Side-channel attacks against TEEs}
\mypar{Page-based attacks} The TEEs that do not manage page tables themselves, such as Intel SGX and ARM SEV, are vulnerable to page-based or controlled-channel attacks. Xu \etal~\cite{xu2015controlled} first demonstrates that an untrusted OS can extract an SGX enclave's secrets by injecting page faults and observing its memory accesses at page granularity. Subsequent work~\cite{BulckWKPS17, WangCPZWBTG17} shows that the enclave page access pattern can also be obtained by observing the access bit in page tables. Many works~\cite{LiZWLC21, LiZLS19, HetzeltB17, Morbitzer0H19, Severed, WernerMAPM19} illustrate attackers can also manipulate page tables to attack AMD SEV/SEVes. However, page-based attacks can only reveal the enclave memory access pattern at a page granularity. 

\mypar{Cache attacks} Cache-based timing attacks improve spatial resolution by exploiting information leakage at a cache line granularity. The majority TEEs still share the cache with the counterpart and allow cache eviction between them, resulting in cache-based attacks. In particular, cache attacks on Intel SGX and ARM TrustZone have been extensively exploited. Most of them probe L1 caches with either frequent interrupts~\cite{MoghimiIE17, HahnelCP17, LapidW18, TruSense, TruSpy, Loadstep} or the Multi-threads~\cite{BrasserMDKCS17, GotzfriedESM17}. Cho \etal~\cite{ ChoZKPL0DA18} exploits both L1 and L2 cache on a single-core and cross-core scenarios, respectively, using prime+count technique in the TrustZone architecture. XLATE~\cite{SchaikGBR18}, ARMageddon~\cite{ARMageddon}, and the attack proposed by Schwarz \etal~\cite{SchwarzWGMM20} probe LLCs. In particular, \cite{SchwarzWGMM20} is an enclave-to-enclave attack. MemJam~\cite{memjam} provides an intra-cache level timing channel and improves the spatial resolution to 4-byte granularity by exploiting the read-after-write false dependencies. 

\mypar{Branch predictor units~(BPUs) attacks} Branch predictor units~(BPUs) have also been exploited to infer the control flow of enclaves. BPU-based attacks either leverage the speculative prediction results made by the BPU or exploit the current state of the BPU. So far, this channel has mainly been explored to attack Intel SGX. For instance, SgxPectre~\cite{chen2018} exploits speculative execution side-channel vulnerabilities in an SGX environment and subverts its security guarantee. The branch shadowing~\cite{SGKKP17} attack demonstrates that enclave-private control flow can be inferred by abusing cache collisions in the CPU-internal branch target Buffer~(BTB). Branchscope~\cite{EvtyushkinRAP18} and Bluethunder~\cite{Bluethunder} exploit the 1-level and 2-level directional predictor in the BPU, respectively. Ryan \etal~\cite{Ryan19} shows that abusing cache collisions in BTB is also effective in attacking ARM TrustZone. 

\mypar{Transient execution attacks} Inspired by the Spectre~\cite{Spectre} and Meltdown~\cite{Meltdown} attacks, more recent attacks leverage the speculative or transient execution to break the confidentiality of SGX enclaves. Typically, the attacker induces transient instructions that access secret data. Execution of transient instructions can modify the state of microarchitectural components based on the secret data. The attacker then probes the microarchitectural component to determine the secret data. The microarchitectural components that have been probed include the store buffer~\cite{Fallout}, load port~\cite{ridl}, line fill buffer~\cite{cacheout,ridl,ZombieLoad}, and interrupt mechanism~\cite{BulckPS18}.


\subsection{Defenses} 
\mypar{Page-based attacks prevention} Defenses against page-based attacks have been extensively studied for Intel SGX. T-SGX~\cite{Shih0KP17} leverages TSX to detect and hide page faults from the untrusted OS. SGX-LAPD~\cite{FuBQL17} proposes to enlarge the page size so as to reduce the information leakage. Shinde \etal~\cite{ShindeCNS16} ensure that all execution traces access the same pages in the same order by adding dummy branches. To avoid page faults, Heisenberg~\cite{Strackxabs} issues dummy instructions to pre-load all enclave pages in a pre-determined order before sensitive code is processed. Autarky~\cite{OrenbachBS20} proposes to modify the Intel SGX architecture so that the enclave no longer reports the faulting page base address to the untrusted OS and does self-paging. CoSMIX~\cite{CoSMIX} and Klotski~\cite{Klotski} adapt enhanced ORAM protocol to randomize the enclave's page access pattern, which also works for protecting other TEEs. D\'{e}j\`{a} Vu~\cite{ChenZRZ17} monitors the enclave execution time with the assistance of TSX. 

\mypar{Cache attacks prevention} To defend against cache-based timing attacks, DR.SGX~\cite{drsgx} and OBFUSCURO~\cite{OBFUSCURO} randomize the data and code with ORAM at cache line granularity. Cloak~\cite{Cloak} and Chen \etal~\cite{ChenLMZLCW18} propose to wrap the secret data and secret handling code with TSX transaction, which is specific to Intel SGX. General cache partition techniques, such as CATalyst~\cite{CATalyst}, page coloring~\cite{COLORIS}, and CacheBar~\cite{ZhouRZ16}, are also effective to protect TEE enclaves from cache-based SCAs.

\mypar{BPUs attacks prevention} Due to the code randomisation, OBFUSCURO can also mitigate the SCAs that exploit BPU, such as Branch shadowing, BranchScope and Bluethunder. To defend against Branch shadowing, in Zigzagger~\cite{SGKKP17}, all conditional and unconditional branches are converted into unconditional branches. Inspired by Zigzagger, in~\cite{MitigatingHLL18} Hosseinzadeh \etal eliminate conditional branches and hide the targets of unconditional branches using a combination of compile-time modifications and run-time code randomization.

\mypar{Generic defenses} Varys~\cite{varys} and Hyperspace~\cite{Hyperspace} are more general defenses that, similar to \sys, mitigate one or two types of SCAs rather than specific attacks. Varys gives the solutions to defend against both interrupt-based and HT-based SCAs, and Hyperspace focuses on HT-based attacks. To detect interrupts, Varys injects code to periodically check a field of the SSA. To detect HT-based attacks, both Varys and Hyperspace propose to run a sibling thread with the enclave thread on the same physical core and do a co-location test periodically. 
To prevent single-stepping enclaves with interrupts, AEX-Notify~\cite{AEX-Notify} propose a hardware ISA extension and a software handler for Intel SGX to prefetch the working set of the next enclave instruction. However, it cannot prevent attacks that require more generic high-resolution probing, such as those exploiting code control flow~\cite{SGKKP17, foreshadow}. 

Compared with the existing software-based defenses, \sys imposes less overhead. Existing defenses need to inject heavy code into the enclave either to obfuscate the memory access patterns or to detect interrupts. \sys mainly needs to modify the enclave's inherent instructions. Moreover, \sys can achieve a stronger security guarantee with less overhead than existing detection-based defenses. \sys just needs to reload the frame base address from the SA more frequently or store/load more generic memory references to/from the SA, yet existing defenses should increase the detection frequency. 

%% file: sections/conclusion.tex
\section{Limitations}

In this section, we discuss the limitations of \sys.

\begin{itemize}[noitemsep, topsep=2pt, partopsep=0pt,leftmargin=0.50cm]
\item[\circled{1}] ~\sys requires a dedicated CPU core to execute the enclave, which introduces resource fragmentation and potentially reduces utilisation. \sys thus introduces a security and performance trade-off where the resources from a dedicated core are used to improve the system resilience against side-channel attacks. 


\item[\circled{2}] ~\sys is not applicable to environments that cannot provide interrupt-free execution on a dedicated CPU core. There are multiple sources of interrupts that are hard to control. 
  Two types of interrupts that are not disabled by tickless mode are (i)~non-maskable interrupts~(NMIs) and (ii)~system management interrupts (SMIs). NMIs are generally used to process critical errors, such as an unrecoverable hardware problem. SMIs can suspend the OS and cause the CPU to go into system management mode~(SMM). The events that trigger SMIs are configured by the firmware during platform initialisation. In general, SMIs are used to offer extended functionality, such as power throttling, thermal management, legacy hardware device emulation, and device wake-up for NICs and USB controllers. When NMIs/SMIs occur, \sys terminates the enclave, causing false positives. To avoid this, \sys requires hardware to remove sources of NMIs/SMIs, such as thermal issues and use of legacy devices. Our experience is that NMIs/SMIs rarely occur in practice.

    

For Intel SGX, when Enclave Page Cache (EPC) memory is not enough, EPC paging also causes AEXs. As Intel has increased EPC memory sizes, EPC paging is no longer an issue~\cite{intelepcsizes}. 
For SGX platforms with limited EPC capacity, \sys can use techniques such as Eleos~\cite{eleos} for user-space virtual memory to eliminate AEXs due to EPC paging. However, when executing SGX enclaves in virtual machines ~(VMs), VM exits cause AEXs. Currently, \sys is not compatible with VMs as it cannot avoid these AEXs. 

   

\item[\circled{3}] With \sys, the attacker cannot keep interrupting the enclave in a single run. However, it could run the enclave multiple times and attack different places. Given enough runs, it is still possible to obtain constant secrets. We can prevent this attack by limiting the attestation frequency on Intel server side. Moreover, we can directly store constant secrets in the SSA, and they will be destroyed by the first interrupt if attacked. 
\end{itemize}

  
\section{Conclusions}

We described \sys, a practical system for protecting TEE enclaves from interrupt-based SCAs. Its key technique is to change the data layout of the enclave in a way that ensures that the untrusted OS executes the enclave in an interrupt-free environment. \sys has been implemented as part of an LLVM-based compiler pass. Our evaluation results show that \sys incurs low overhead while achieving a strong security guarantee.
